\documentclass[journal=jacsat,manuscript=article]{achemso}

\usepackage{chemformula} 
\usepackage[T1]{fontenc} 
\usepackage{caption}
\usepackage{graphicx}
\usepackage{hyperref}
\DeclareCaptionLabelFormat{adja-page}{\\}

\author{Aravind P. Anthur}
\altaffiliation{Equal contribution. [Version submitted to Nano Letters. \\ Revised version: \url{https://doi.org/10.1021/acs.nanolett.0c03601}]}
\author{Haizhong Zhang}
\altaffiliation{Equal contribution. [Version submitted to Nano Letters. \\ Revised version: \url{https://doi.org/10.1021/acs.nanolett.0c03601}]}
\author{Ramon Paniagua-Dominguez}
\altaffiliation{Equal contribution. [Version submitted to Nano Letters. \\ Revised version: \url{https://doi.org/10.1021/acs.nanolett.0c03601}]}
\affiliation[IMRE, A*STAR]{Institute of Materials Research and Engineering, A*STAR (Agency for Science, Technology and Research) Research Entities, 2 Fusionopolis Way, $\#$08-03 Innovis, Singapore 138634.}

\author{Dmitry Kalashnikov}
\author{Son Tung Ha}
\author{Tobias Wilhelm Wolfgang Mass}
\author{Arseniy I. Kuznetsov}
\email{arseniy_kuznetsov@imre.a-star.edu.sg}
\author{Leonid Krivitsky}
\affiliation[IMRE, A*STAR]{Institute of Materials Research and Engineering, A*STAR (Agency for Science, Technology and Research) Research Entities, 2 Fusionopolis Way, $\#$08-03 Innovis, Singapore 138634.}
\email{leonid_krivitskiy@imre.a-star.edu.sg}

\title{Continuous wave second harmonic generation enabled by quasi-bound-states in the continuum on gallium phosphide metasurfaces}

\abbreviations{SHG, BIC}

\begin{document}


\begin{abstract}
Resonant metasurfaces are an attractive platform for enhancing the non-linear optical processes, such as second harmonic generation (SHG), since they can generate very large local electromagnetic fields while relaxing the phase-matching requirements. Here, we take this platform a step closer to the practical applications by demonstrating visible range, continuous wave (CW) SHG. We do so by combining the attractive material properties of gallium phosphide with engineered, high quality-factor photonic modes enabled by bound states in the continuum. For the optimum case, we obtain efficiencies around 5e-5 $\%$ W$^{-1}$ when the system is pumped at 1200 nm wavelength with CW intensities of 1 kW/cm$^2$. Moreover, we measure external efficiencies as high as 0.1 $\%$ W$^{-1}$ with pump intensities of only 10 MW/cm$^2$ for pulsed irradiation. This efficiency is higher than the values previously reported for dielectric metasurfaces, but achieved here with pump intensities that are two orders of magnitude lower.  
\end{abstract}

Nonlinear optical processes like second harmonic generation (SHG) has attracted the interests of researchers for more than sixty years  \cite{Boyd_NLO, Rottwitt_NLO:PandA}. SHG is utilized for a broad range of applications including lasers, quantum optics, spectroscopy, imaging and pulse-width measurement, to name a few \cite{Armstrong_1967, Heinz_1982, Galli2019, Stolzenburg2010, Freund_1986}. One of the fundamental goals for any practical use of SHG is to achieve efficient conversion of the fundamental frequency pump to the second harmonic signal. Several materials and device designs have been proposed and demonstrated to enhance the efficiency. These, however, come at the price of several constraints related to the operational wavelength ranges, necessary pump intensities, phase-matching requirements and/or the size of these devices. Towards reducing the device dimensions and overcoming the phase matching constraints, there are ongoing efforts in utilizing metasurfaces, arrays of nanostructures that exploit resonant enhancement of the pump \cite{Kildishev2013, Lapine2014, Genevet2017, Krasnok2017, Staude2019}. Using this concept, high efficiency has been achieved, for example, in plasmonic metasurfaces \cite{DeCeglia2015, Celebrano2015, Canfield2007}. While leading to very high local field, plasmonic metasurfaces have the disadvantage of having intrinsic dissipative losses, which result in relatively low damage thresholds for this kind of devices. As an alternative, dielectric and semiconductor metasurfaces have recently emerged as a promising platform \cite{Arseniy2016, Kruk2017b, Ramon2019}. They offer lower intrinsic material losses and higher damage thresholds, as well as, potentially, high non-linear coefficients. Using materials that lack the inversion symmetry of the crystal lattice allows SHG in the volume, rather than at the surface (as is the case of metals). In this regard, semiconductor metasurfaces with multiple quantum wells are among those with the largest non-linear coefficients,  and have been used to achieve high conversion efficiencies \cite{Lee2016, Sarma2019}. The main limitation of this approach, however, is that the SHG signal in these structures saturates at relatively low pump intensities, limiting the total amount of power that can be converted in the device. Resonant dielectric and semiconductor metasurfaces using volumetric SHG, and having higher damage and saturation thresholds, have been used to demonstrate very high SHG efficiencies utilizing cleverly engineered optical modes \cite{Carletti2015, Gili2016, Camacho-Morales2016, Cambiasso2017, Li2020, Koshelev2020}. 

In this regard, a novel design strategy of dielectric nanoantennae and metasurfaces based on the concept of bound-states in the continuum (BIC) have recently attracted interest in the community \cite{Koshelev2020, Hsu2016, Kodigala2017, Tony2018, Kupriianov2019, Diego2019, Mengfei2020, Vasilii2020}. The main reason is that it allows to achieve very high quality factor (Q) modes, which can be easily tuned by varying the device geometry, enabling highly efficient nonlinear optical processes \cite{Xu2019, Zhou2020, Koshelev2020, Bernhardt2020}. Despite the advances in the optical mode and material engineering, all nanoantenna- and metasurface-based SHG devices demonstrated so far have been realized in the high-intensity, pulsed regime (typically involving femtosecond lasers), which hinders their use in many practical applications that work in the low intensity CW regime. This also creates an additional limitation due to a limited overlap of the short-pulse laser spectrum with the narrow-band resonances of high-Q metasurface, which further limits the conversion efficiency \cite{Shcherbakov2019}. In this work, we combine an emerging material platform for non-linear applications, namely gallium phosphide (GaP) \cite{Grinblat2020, shcherbakov2020}, and a metasurface design supporting a high-Q quasi-BIC mode that exploits the symmetries of the non-linear tensor, to generate continuous wave (CW) SHG at low pumping intensities with reasonable conversion efficiencies. We also show that, in the pulsed regime, we achieve external efficiencies that are higher than the results reported in the literature for dielectric metasurfaces to the best of our knowledge, despite using two orders of magnitude lower pumping intensities. 

\begin{figure}[t]
\centering
\includegraphics[scale=0.63]{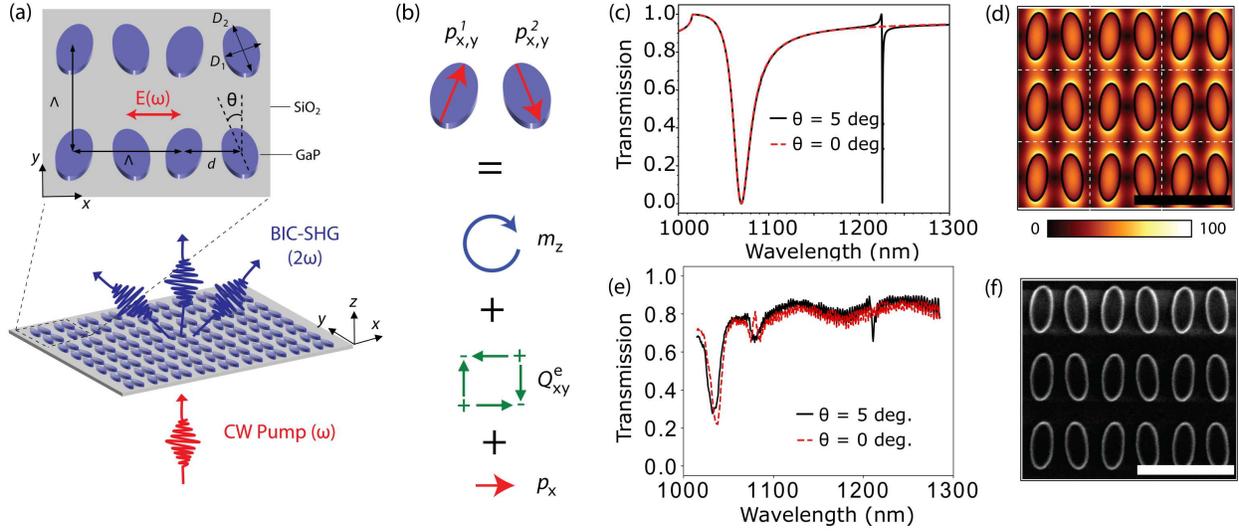}
\caption{(a) Schematic of the metasurface, comprising a square lattice (periods, $\Lambda$ = 700 nm) of dimers formed by two elliptical cylinders (with minor and major axes D$_1$ and D$_2$, respectively, height of 150 nm and center-to-center separation of d = 340 nm) with a relative tilting by an angle $\theta$ between their major axes. (b) Schematic representation of origin of the quasi-BIC. A single in-plane electric dipole is induced in each particle in the dimer, containing x- and y-components ($p_{(x,y)}^i$). These, when summed up, give origin to an out-of-plane magnetic dipole ($m_z$) and in-plane electric quadrupole (Q$_{xy}^e$), and an x-oriented electric dipole ($p_x$), whose amplitude can be controlled by $\theta$. Of these, only $p_x$ can couple to normally incident, x-polarized plane waves. (c) Simulated transmission spectra of x-polarized normally incident plane waves traveling through the metasurface with $\theta=0^{\circ}$ (red) and $\theta=5^{\circ}$ (black). (d) Amplitude of the electric near-field distribution at the quasi-BIC resonance in the xy-plane passing through the center of the particles. (e) Measured transmission spectra of x-polarized, normally incident light traveling through the fabricated metasurfaces with $\theta=0^{\circ}$ (red) and $\theta=5^{\circ}$ (black). (f) SEM image of the fabricated sample with $\theta=5^{\circ}$ with scale bar of 1 um.}
\label{Figure1}
\end{figure}

Our metasurface consists of a square array of nano-dimers comprising two GaP cylinders with elliptical cross section,  as shown in Fig. \ref{Figure1}(a). They have period of $\Lambda$=700 nm, minor axis length of D$_1$=200 nm, major axis length of D$_2$=460 nm, height of 150 nm and center-to-center distance of d=340 nm. When the major axes of these cylinders are perfectly parallel to each other,the system supports a perfect BIC at the gamma point for the selected pump wavelength ($\lambda$ $\approx$ 1225 nm) \cite{Joannopoulos, Koshelev2018}. This BIC stems from the excitation of an in-plane electric quadrupole mode, Q$_{xy}^e$, and an out-of-plane magnetic dipole mode, m$_z$, in each dimer, which in turn result from individual electric dipoles excited in each nanocylider, as shown in Fig. \ref{Figure1}(b). Due to the symmetry of the waves radiated from these modes, they cannot couple to plane waves emerging from the system at normal incidence and the system becomes decoupled from the radiation continuum. A simple way to visualize the emergence of this BIC is the following. Each unit cell, comprising a dimer, radiates according to the sum of the in-plane electric quadrupole and the out-of-plane magnetic dipole. Since each of these two modes has vanishing radiation intensity in the direction normal to the metasurface, their sum obviously also has so. Now, when all the unit-cells oscillate in phase, a situation corresponding to the gamma point, and when the lattice period is sub-diffractive, this normal direction is the only direction in which the system is allowed to radiate. However, when radiation from each unit cell vanishes in the normal direction, as is the case of these dimers, overall radiation is forbidden from the system and the modes become bounded to the system (hence the name BIC). By reciprocity, this also implies that no energy can be coupled into these modes from normally incident plane waves. In order to be able to do so, a certain radiation leakage channel is necessary. In this system, an easy way to open such a channel is by symmetry breaking, realized via tilting of the ellipses in opposite directions around the z-axis by a certain angle $\theta$, as shown in Fig. \ref{Figure1}(a). In such situation, the system might support an additional in-plane, x-oriented electric dipole moment, p$_x$ (see Fig. \ref{Figure1}(b)) that opens the possibility to couple power in and out of the system, when illuminated by a normally-incident x-polarized plane wave (see Fig. \ref{Figure1}(a)). This effectively transforms the perfect BIC, characterized in the lossless case for having Q $\to \infty$, into the quasi-BIC, with large but finite Q. The amplitude of p$_x$, which can be controlled by the angle $\theta$, determines the amount of leakage and, thus, the Q of the mode \cite{Koshelev2018}.

We use this quasi-BIC mode to couple power into the system and achieve enhancement of the pump intensity, thereby achieving highly efficient SHG. Figure \ref{Figure1}(c) shows the simulated transmission spectra for two different metasurfaces, one supporting a perfect BIC ($\theta = 0^{\circ}$) and one supporting a quasi-BIC ($\theta=5^{\circ}$). As can be seen, a spectrally narrow resonance at $\sim$ 1225 nm can be observed for the case of $\theta=5^{\circ}$, corresponding to the excitation of the quasi-BIC. The Q factor is estimated to be more than 4000, obtained through a fitting of the curve to the Fano formula \cite{Fano1961}. As expected, this dip is not present for the case in which the perfect BIC is formed ($\theta=0^{\circ}$), as the mode becomes uncoupled from external radiation and, thus, cannot be excited by the pump. In Fig. S1, we plot the evolution of this mode as a function of  $\theta$, from which the narrowing and amplitude decrease of the resonance as the angle reduces becomes clearly apparent, as expected for a mode evolving towards a perfect BIC. Note that this behaviour is not observed for the broad resonance excited at shorter wavelengths, which has its origin in a simple, in plane electric dipole ($p_x$) that is almost insensitive to the tilting angle. Figure \ref{Figure1}(d) shows the simulated enhancement of the electric near-field amplitude at the quasi-BIC excitation condition. The result indicates that inside the structures, field enhancements above 50 can be obtained, corresponding to intensity enhancement of more than 2500. While, theoretically, arbitrarily high values of Q factors are possible by simply reducing $\theta$, in practice they are limited by the size of the array, the material losses and any fabrication imperfections. In our experiments, the minimum angle for which we observed the quasi-BIC mode is $\theta=5^{\circ}$, and is therefore the case presented here. A detailed analysis of the multipolar moments supported by this system is presented in Fig. S2, corroborating the interpretation of the quasi-BIC origin given above.

We choose GaP for fabricating our devices because it combines all key features that make an attractive platform for non-linear processes at the nanoscale. First, it has a large second-order nonlinear coefficient (d$_{36}$ $\approx$ 70 pm/V for $\lambda \approx$ 1000 nm) \cite{Shoji_1997}. Second, it has a relatively large band gap that pushes the cut-off wavelength down to $\lambda \approx $ 550 nm, together with a high refractive index, n $\approx $ 3 (for $\lambda\approx$ 1200 nm) \cite{Bond1965}. And last but not the least, nanofabrication of GaP is well established in the semiconductor industry \cite{Wilson2019}. Since the GaP crystal has zinc-blende structure, it is critical to know and utilize the nonlinear tensor appropriately to achieve high SHG efficiency. For that, we first analyze the crystal axis orientation of our wafer, which turns out to be tilted with respect to the surface normal \cite{Anthur2020}. Then we fabricate our metasurfaces with an orientation that maximizes the SHG. We do so using electron beam lithography (EBL) followed by reactive ion etching (RIE). Both crystal orientation and fabrication details are provided in the supporting information. The fabricated samples are of high quality, as indicated by the good agreement between the measured transmission spectra (Fig. \ref{Figure1}(e)) with the simulated ones (Fig. \ref{Figure1}(c)), as well as the scanning electron microscopy (SEM) images of the sample (see Fig. \ref{Figure1}(f) for the $\theta=5^{\circ}$ case and Fig. S3 for all the cases measured).  Despite the slight blueshift in the measured position of the transmission dip with respect to the simulated one, both spectra are in good agreement, obtaining a quality factor of approximately 2000 for the experimental quasi-BIC structure.  For these transmission measurements, the sample was illuminated using an x-polarized broad band halogen light source (see supporting information).

\begin{figure}[t]
\centering
\includegraphics[scale=0.7]{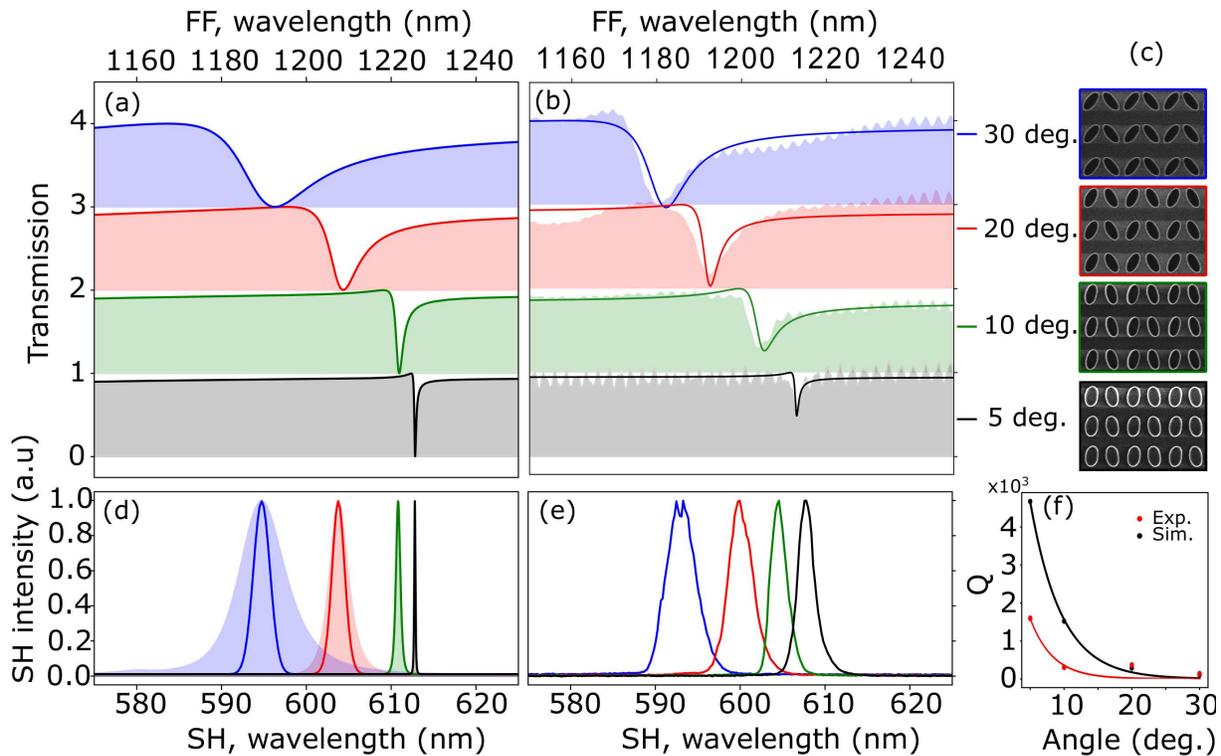}
\caption{(a) Simulated and (b) measured transmission spectra near the quasi-BIC resonance of x-polarized, normally incident light traveling through metasurfaces with varying  $\theta$, from 30$^{\circ}$ to 5$^{\circ}$. (c) SEM images of the fabricated samples. (d) Simulated and (e) measured SHG spectra when the metasurfaces in (a) and (b) are illuminated by a pulsed pump (FWHM around 5 nm) centered at the wavelength giving maximum SHG (solid lines). In (d), we also show as shaded area the SHG spectra computed assuming an infinitely narrow pump, whose frequency is scanned in the range shown. (f) Quality factor (Q) retrieved from simulated and measured transmission spectra through a fitting to the Fano formula.}
\label{Figure2}
\end{figure}

The simulated and measured transmission spectra of metasurfaces with different angle $\theta$, spanning from $5^{\circ}$ to $30^{\circ}$, are given in Figs. \ref{Figure2}(a) and \ref{Figure2}(b), respectively, with the corresponding SEM images given in Fig. \ref{Figure2}(c). As expected, the Q factor experiences a dramatic increase as $\theta$ reduces, while the resonance wavelength redshifts. Simulations and experiments are in a very good agreement with respect to the relative spectral position of the resonance for different cases and the Q factors (see Fig. \ref{Figure2}(f) for a direct comparison). The overall slight blueshift ($\sim$20 nm) of the experimental spectra with respect to the theoretical ones can be attributed to a slight deviation in the height of the structures. The quasi-BIC resonant wavelengths are designed to be around 1200 nm so that the SHG wavelength is above the cut-off wavelength of the GaP. It should be noted that all the spectra are measured under x-polarized illumination, and that no resonances are observed for y-polarized light.

We now use the near field enhancement inside the structures enabled by the quasi-BIC resonances to obtain SHG and study its efficiency as a function of the angle $\theta$. We first pump the metasurfaces using a pulsed laser from an optical parametric oscillator (APE) with a pulse duration of around 200 fs and wavelength that is tunable from 1100 nm to 1400 nm. The light is focused onto the sample with a plano-convex lens with a focal length of 25.4 mm. The forward generated second harmonic light is collected using an objective lens with a numerical aperture (NA) of 0.95 in air. The SHG signal is filtered using a bandpass filter and analyzed (see supporting information and Fig. S4 for more details on the experimental system). 

In Figs. \ref{Figure2}(d) and \ref{Figure2}(e), we show as solid lines, respectively, the simulated and measured SHG spectra obtained when the central wavelength of the pulsed pump is tuned to give the maximum SH signal. In our simulations, we assume that the pump pulse has a Gaussian shape, with 5 nm full-width at half maximum (FWHM). In Fig. \ref{Figure2}(d), we also show as a shaded area the simulated SHG assuming a single frequency pump, scanned over the range of wavelengths. All simulated results are computed in two steps. First, the electric field distribution (\textbf{E}($\omega_{FF}$)) inside the nanostructures at the fundamental frequency ($\omega_{FF}$) is computed. Then, it is used to calculate the non-linear polarization density, which serves as the source for a second simulation at the second harmonic frequency, using the relation \textbf{P}(2$\omega_{FF}$)=2$\epsilon_0\overline{\overline{\chi}}^{(2)}$ \textbf{E}($\omega_{FF}$). Here, $\epsilon_0$ is the vacuum permittivity and $\overline{\overline{\chi}}^{(2)}$ is the appropriate non-linear susceptibility tensor in the laboratory frame, which takes into account the crystal axis orientation in the experiment, as explained in the supporting information. Figure \ref{Figure2}(e) is measured using a spectrometer (Ocean Optics, USB4000)  with a spectral resolution of  $\sim$1 nm which explains why the SH spectra of the 10$^{\circ}$ and 5$^{\circ}$ have similar spectral widths.  From Figs. \ref{Figure2}(d) and (e) one can see that the wavelength of maximum SHG signal exactly corresponds to the quasi-BIC condition. Moreover, a noticeable narrowing of the SHG spectra is observed when the pump is tuned to the resonant wavelength of the metasurface; its FWHM evolving from that of the pump to being determined by the resonance itself. In the experimental data, this narrowing saturates when it hits the resolution limit of our spectrometer. 

\begin{figure}
\centering
\includegraphics[scale=0.54]{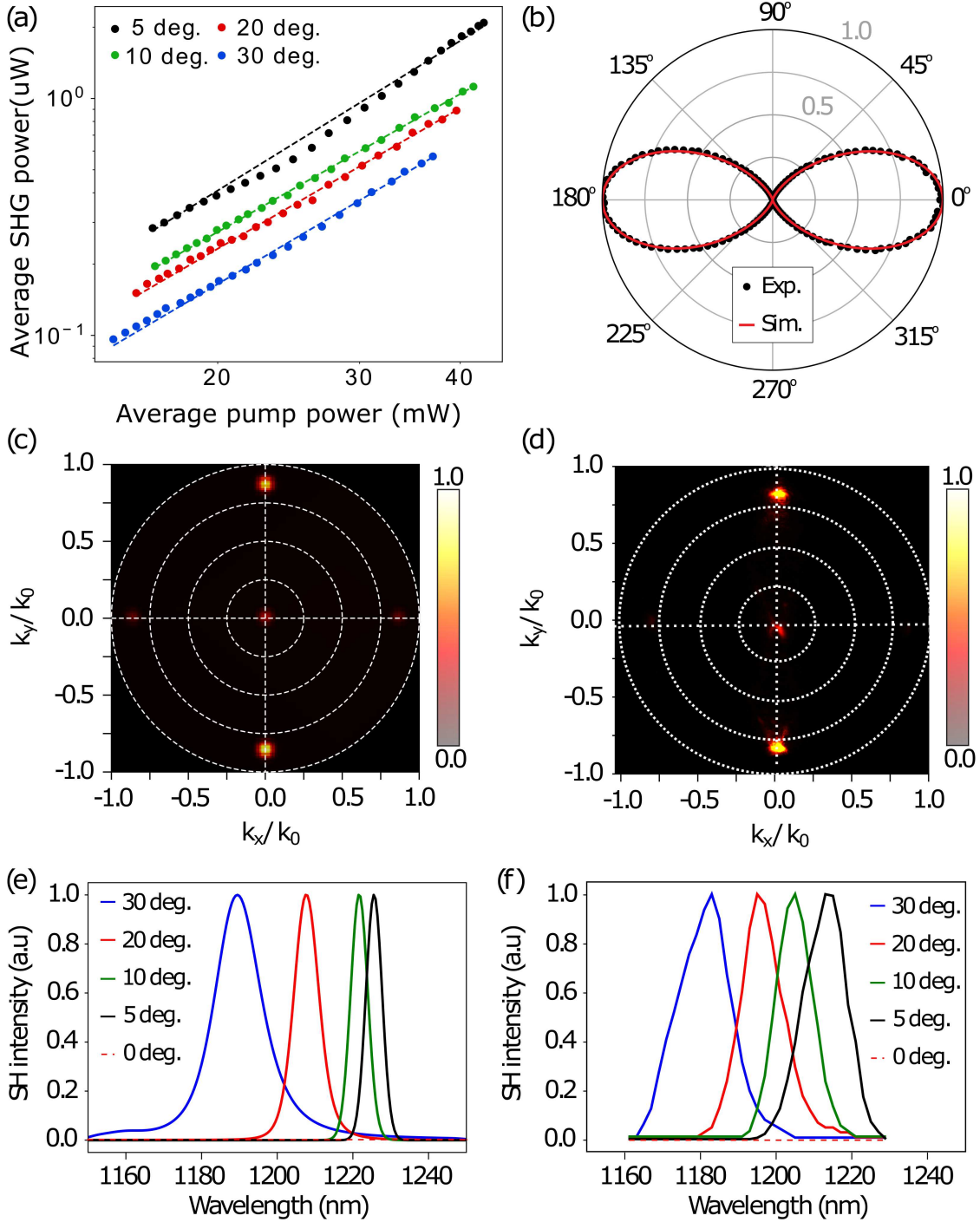}
\caption[]{(a) Measured SHG power as a function of the average pulsed pump power. For each angle, the pulsed central wavelength is tuned to the wavelength giving the maximum SHG. Fits to the equation $P_{SHG}=aP_{pump}^b$, are shown as dotted lines, while experimental points are shown as circles. The exponent values obtained from the fits are 2, 1.97, 1.93, and 2.1  for 30$^{\circ}$, 20$^{\circ}$, 10$^{\circ}$ and 5$^{\circ}$ respectively. The R-squared values are $\sim$ 1.  (b) Simulated (red solid line) and measured (black circles) SHG power as a function of the pump polarization (0$^{\circ}$ corresponding to the polarization parallel to the x-axis, i.e. with the axis of the dimer) for the metasurface with $\theta = 5^{\circ}$. (c) Simulated and (d) measured SHG back-focal plane}
\end{figure}

\begin{figure}
    \centering
    \captionsetup{labelformat=adja-page}
    \ContinuedFloat
    \caption{images at the maximum of SHG for the metasurface with $\theta=5^{\circ}$. (e) Simulated and (f) measured SHG spectra when the central frequency of the pulsed pump is scanned around the quasi-BIC condition.}
    \label{Figure3}
\end{figure}

Figure \ref{Figure3}(a) shows the measured SHG power in the forward direction as a function of the input pump power for the four cases shown in Fig. \ref{Figure2}(e). The input pump power is varied using a variable neutral density filter at the input. The SHG power is measured using a power meter for the pulsed pump and using a lock-in amplifier system with an amplified photo detector (APD) for the CW pump (see supporting information and Fig. S4 for details). Fitting the curve to a function with the shape $P_{SHG}=aP_{pump}^b$, we found the nearly quadratic dependence expected for SHG. As can be seen from Fig. \ref{Figure3}(a), the average collected SHG power for $\theta=5^{\circ}$ is around 1.44 $\mu$W for 37 mW of average input pump power. This corresponds to an external SHG efficiency of 4e-5 ($P_{SHG}$/$P_{pump}$) and 0.1 $\%$ W$^{-1}$ ($P_{SHG}$/$P_{pump}^2$), for a pump intensity of approximately 10 MW/cm$^2$. This level of efficiency, to the best of our knowledge, is better than previously reported results in the literature, despite being obtained here using 100 times lower intensities (see Table S1). It is worth noting that, for this input power, the SHG power is not saturated. We also note the evident impact of the narrowing of the quasi-BIC resonance on the SHG as the angle varies from $\theta=30^{\circ}$ to $\theta=5^{\circ}$. As expected, the increase in Q factor has an associated increase of the SHG, reaching values for the $\theta=5^{\circ}$ that is 2.5 times larger than the $\theta=30^{\circ}$ case. This value is lower than what would be expected if one simply considers the Q factor increase. This is attributed to the fact that the width of resonance becomes significantly narrower than the pump at small angles ($\theta$), and thus only a part of the femtosecond pump spectra is used for the generation of the SH signal. This finally leads to a limited enhancement in the SHG for the 5 degree compared to the 30 degree and inspires us to pursue CW SHG in this system.  

Before that, we corroborate the quasi-BIC origin of the observed SHG by analyzing its dependence on the polarization of the pump. As mentioned before, the quasi-BIC resonance can only be excited when the normally incident wave has its electric field polarized along the axis of the dimer (x-axis). The perpendicular, y-oriented polarization, cannot couple to the electric quadrupole Q$_{xy}^e$, nor to the out-of-plane magnetic dipole mode, m$_z$, as it follows from the symmetry considerations. Therefore, we expect significant SHG only when the incident electric field is oriented along the x-axis. In the experiment, the polarization of the pump is rotated using a half wave plate and the power of the SHG at the output is measured without any polarizer (either at the input or the output of the sample). The measurement results are plotted as circles in Fig. \ref{Figure3}(b), together with the simulation predictions, represented as solid lines. As expected, the SHG power is highest for x-polarized pump, while for the fully y-polarized pump it is below the detection limit of our measurement system, involving amplified photodetector and lock-in amplifier. For completeness, the simulated and measured back-focal plane images of the far-field SHG radiation pattern in the forward direction are given in Figs. \ref{Figure3}(c) and \ref{Figure3}(d), respectively, for the $\theta=5^{\circ}$ sample. The measured back-focal plane image uses an objective with an NA of 0.95 to collect the SHG signal. As can be seen from Fig. \ref{Figure3}(d), five spots are observed, corresponding to the direct forward emission and the SHG coupling into the four diffraction orders opened at the SH frequency, 2$\omega_{FF}$. Two of the five spots are more intense than the other three, in good agreement with the simulations (Fig. \ref{Figure3}(c)). Since the crystal axis is rotated by an angle of 15$^{\circ}$ with respect to the normal to the wafer, we observe the beam along the direction of propagation at 0$^{\circ}$ angle, which would otherwise be zero due to the nonlinear tensor for GaP if the axis of the crystal would be parallel to the normal. 

Finally, we sweep the central frequency of the pulsed pump in a range of wavelengths around the (quasi-)BIC for all samples, from $\theta=30^{\circ}$ to $\theta = 0^{\circ}$, and plot the simulated and measured normalized SHG intensity spectra in Figs. \ref{Figure3}(e) and \ref{Figure3}(f), respectively. As can be seen, maximum SHG is obtained when the central wavelength corresponds to the resonant dip observed in the transmission spectra of Figs. \ref{Figure2}(a) and \ref{Figure2}(b). Also, as expected, the FWHM of the curves gets narrower for decreasing angle, but becomes limited by the convolution with pump pulse width. In fact, as mentioned before, the mismatch between the (narrower) spectral width of the quasi-BIC resonance for small angles and the (broader) pulsed pump implies that part of the incident power is not coupled to the mode and, therefore, to the SHG process (the quasi-BIC acting effectively as a filter). Thus, we move on to study the samples under narrow-band CW pump instead, for which the whole input power can be efficiently used in the SHG.

\begin{figure}
    \centering
    \includegraphics[scale=0.45]{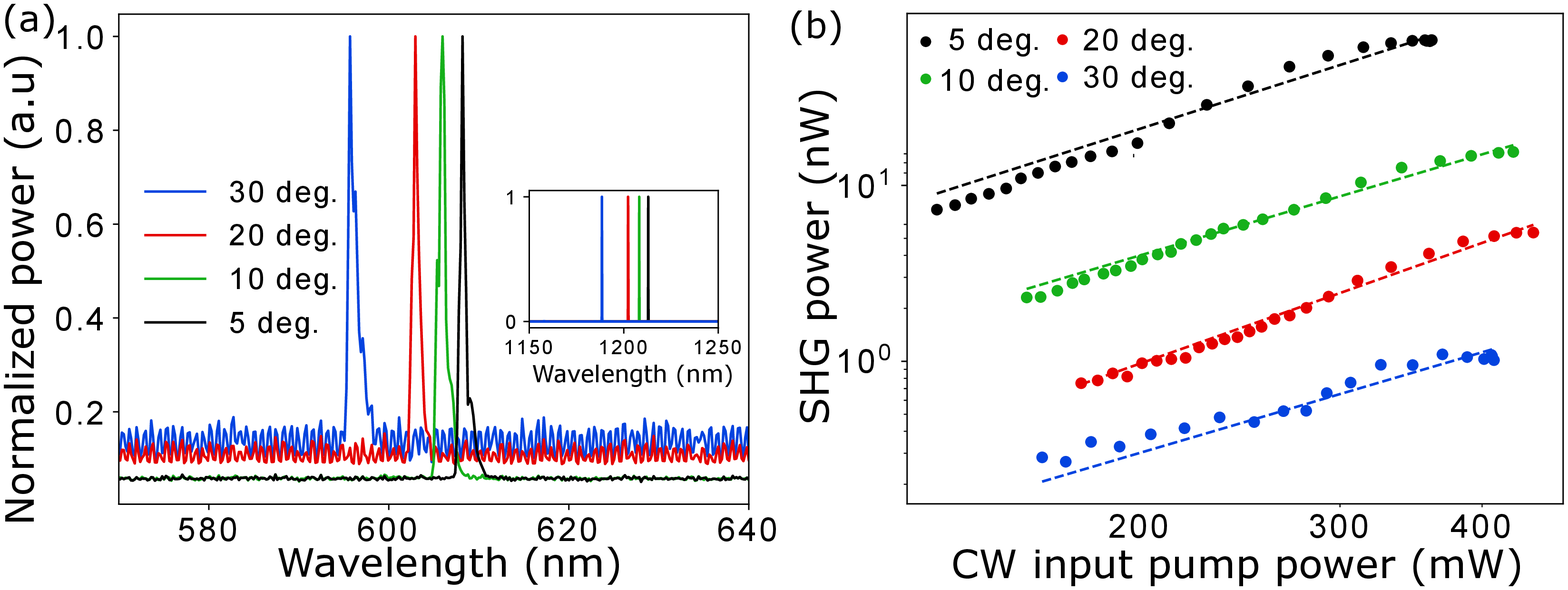}
    \caption{(a) Spectra of the SHG when the CW pump is tuned to the resonance of  30$^{\circ}$, 20$^{\circ}$, 10$^{\circ}$ and 5$^{\circ}$ samples, and the corresponding pump spectra given in the inset. (b) The SHG power as a function of input CW pump power for various angles of the dimer. Fits to the equation $P_{SHG}=aP_{pump}^b$, are shown as dotted lines, while experimental points are shown as circles. The exponent values obtained from the fits are 1.9, 2.3, 1.91, and 2.1  for 30$^{\circ}$, 20$^{\circ}$, 10$^{\circ}$ and 5$^{\circ}$ respectively. The R-squared values for the fit are $\sim$1.}
    \label{Figure4}
\end{figure}

In Fig. \ref{Figure4}(a), we show the optical spectra of the maximum SHG obtained when the different samples are excited with a tunable CW laser (see supporting information for details). As an inset in that figure, we show the spectra of the pump used to obtain the SHG. We note that maximum SHG is observed when the CW pump is tuned to the corresponding resonant transmission dip of each of the metasurface arrays. Notably, we observed and measured SHG using $\sim$ kW/cm$^2$ levels of intensity in all the arrays, with angles ranging from $\theta=30^{\circ}$ to $\theta=5^{\circ}$. As before, SHG is not observed for the case of parallel ellipses, i.e. for $\theta=0^{\circ}$. 

The variation of the SHG power as a function of the input power of the CW pump is shown in Fig. \ref{Figure4}(b). As expected, one can observe an increase in the SHG power (and thus the efficiency) as $\theta$ decreases from $30^{\circ}$ to $5^{\circ}$. In our experiments, the maximum SHG power is observed for the $\theta=5^{\circ}$ case, reaching values around 70 nW for 360 mW of input pump power. This translates into an external SHG efficiency of 2e-7 ($P_{SHG}$/$P_{pump}$) and 5e-5 $\%$ W$^{-1}$ ($P_{SHG}$/$P_{pump}^2$), for a pump intensity as low as 1 kW/cm$^2$. As compared to the pulsed case, since the CW linewidth is less than 1 MHz, the narrowing of the resonance (or increase of the Q factor) does not cut any incident power, and thus the enhancement is not limited when $\theta$ decreases. Therefore, in Fig. $\ref{Figure4}$, the impact of the increase in Q factor is more clearly observed in the SHG, which is almost two orders of magnitude larger for $\theta$=5$^{\circ}$ than it is for $\theta=30^{\circ}$, compared to only 2.5 times increase in the pulsed case.

The generation of the SH using CW and pulsed pump also shows that the metasurfaces utilizing quasi-BIC based resonance enhancement of the pump fields can be used for applications with different requirements by tuning the symmetry-breaking of these structures thereby achieving precise tuning of the Q factors. For example, for pulse width measurements in auto-correlators, the spectral width of the resonance of these metasurfaces need to be larger than the spectral width of the laser. For applications where CW pumps are used, the asymmetry can be tuned for highest Q factor and conversion efficiency. Thus these structures have the flexibility to be utilized in various applications by controlling the device geometry. 

\section*{Conclusion}
In conclusion, we have demonstrated high efficiency second harmonic generation (SHG) from GaP metasurfaces with sub-wavelength thicknesses ($\sim\lambda$/8). This is achieved by combining the emerging material platform of GaP, offering high non-linearity, high refractive index, low loss and ease of fabrication, with high-Q optical resonances enabled by quasi bound-states in the continuum (BIC). We achieve these quasi-BIC resonances by breaking the metasurface unit-cell symmetry, that controllably opens a leaky channel in the normal direction. Opening of leaky channel allows the coupling of radiation in-and-out of the otherwise perfect BIC, formed by in-plane electric quadrupole and out-of-plane magnetic dipole modes. We observe a dramatic increase of the Q factor of the resonance as the asymmetry parameter decreases, which translates into increased SHG efficiency, $P_{SHG}$/$P_{pump}^2$ ($P_{SHG}$/$P_{pump}$), reaching values of 0.1 $\%$ W$^{-1}$ (4e-5) for pulsed pumps with intensities of approximately 10 MW/cm$^2$. This level of external efficiency is obtained without reaching saturation and is higher than the values previously reported in the literature, to the best of our knowledge, but achieved here using 100 times lower intensities. We also demonstrated SHG in these metasurfaces using a continuous wave (CW) pump. We generated CW SHG using pump with intensities of 1 kW/cm$^2$, and achieved conversion efficiencies of 5e-5 $\%$ W$^{-1}$ (2e-7) in the metasurface with the highest Q factor. We believe that our work shows that the combination of suitable material platforms and properly engineered optical resonances might push SHG in metasurfaces to levels closer to those required for practical applications, including those that demand continuous wave, second harmonic generation.



\begin{acknowledgement}

We acknowledge the support of the Quantum Technologies for Engineering (QTE) program of A*STAR, IET A F Harvey Engineering Research Prize 2016 and A*STAR SERC Pharos program, Grant No. 152 73 00025 (Singapore).

\end{acknowledgement}

\subsection*{Author contributions}
R. P.-D. proposed and simulated the structures. H. Z. Z. fabricated the samples. A. P. A. constructed and performed the optical characterization measurements. S. T. H. performed the transmission measurements. T. W. W. M. simulated the back-focal plane images. D. K. helped in the optical characterization. A. P. A wrote the first draft of the manuscript, with inputs from R. P.-D. and H. Z. Z. A. I. K. and L. K. conceived the idea and supervised the project. All authors contributed to the final version of the manuscript

\subsection*{Competing interests}
The authors declare no competing financial interest.

\begin{suppinfo}

The following files are available free of charge.
\begin{itemize}
  \item Supplementary material contain the details of simulation, fabrication and experimental measurement system. Supplementary material also contains the following figures and table. Colour map representing the metasurface transmission as a function of wavelength and angle of tiling ($\theta$) of the elliptical cylinders forming the dimers in the metasurface. Multipolar decomposition of one unit-cell of the metasurface supporting a quasi-BIC with $\theta = 5^{\circ}$. Scanning electron microscope images of the different fabricated metasurfaces. Schematic of the experimental setup. Comparison of the state-of-the-art performances of SHG in dielectric metasurfaces.
\end{itemize}

\end{suppinfo}

\clearpage

\section*{Supplementary material: continuous wave second harmonic generation enabled by quasi-bound-states in the continuum on gallium phosphide metasurfaces}





\setcounter{figure}{0}
\renewcommand{\thefigure}{S\arabic{figure}}

\setcounter{table}{0}
\renewcommand{\thetable}{S\arabic{table}}

\subsection*{Simulations}

Simulations of SHG were carried out using the Finite Element Method in a commercial software package (COMSOL Multiphysics). The simulations consist of two steps. 
In the first one, the transmission spectra and the near-fields inside the nanostructures comprising the metasurface, \textbf{E}($\omega_{FF}$), are computed at the fundamental frequency, $\omega_{FF}$. For that, we simulate the infinite periodic system by using Periodic Boundary Conditions in the x- and y-directions (transverse to the incident wave propagation). Periodic ports are used at the top and bottom to excite the system and record the transmitted and reflected fields. The GaP material parameters are taken from the software database, which in turn are taken from the literature \cite{Bond1965}. A homogeneous surrounding environment with refractive index 1.46 is taken. In the second step, the near-fields recorded in the first one are used to compute the non-linear dipole moment density, which is used as the excitation source in the simulations. This non-linear dipole moment density is computed through the expression:

\begin{equation}
    \mbox{\textbf{P}}(2\omega_{FF}) = 2\epsilon_0 \overline{\overline{\chi}}^{(2)}\mbox{\textbf{E}}(\omega_{FF}),
    \label{eq1}
\end{equation}

where $\epsilon_0$ is the vacuum permittivity and $\overline{\overline{\chi}}^{(2)}$ is the non-linear susceptibility tensor of the crystal in the laboratory frame. To obtain the correct results, one should note that, in our GaP wafers, the material is grown such that the crystal axis subtends an angle of 15$^{\circ}$ with respect to the normal to the wafer surface. Thus, we first need to express the near-fields, as obtained in the first simulation, in the crystal frame, in which the non-linear susceptibility tensor has only off-diagonal components and they are all equal (d$_{14}$  = d$_{25}$  = d$_{36}$ $\approx$ 70 pm/V) \cite{Shoji_1997}. This can be achieved simply using the appropriate rotation matrix. Once the fields in the crystal frame are known, we compute the non-linear dipole moment in this frame using expression in Eq. \ref{eq1}. This is brought back again to the laboratory frame by the inverse rotation matrix, to perform the simulations. In the simulation, we use again Periodic Boundary Conditions in the transverse directions but substitute the Periodic Ports by Perfectly Matched Layers in the z-directions (top and bottom of the simulation). For the computation of the far-field pattern given in Fig. 3(c), a homemade implementation of Stratton-Chu formulas is used \cite{Stratton1941}. For that, the complex electromagnetic fields in one unit-cell are recorded in a plane at distance of 500 nm from the structures. These are used to generate a finite array of 50 x 50 periods, which is then rigorously propagated to the far-field.

\subsection*{Fabrication}
The GaP active layer ($\sim$400 nm) is first grown on a gallium arsenide (GaAs) substrate with an AlGaInP buffer layer by metal-organic chemical vapor deposition (MOCVD) to reduce the lattice mismatch between the GaAs substrate and GaP layer. Then this structure is directly bonded to the sapphire substrate after depositing $\sim$SiO$_2$ layer on top of both the surfaces. The AlGaInP/GaAs substrate is then removed by wet etching. The fabrication of the GaP nanostructures start with a standard wafer cleaning procedure (acetone, iso-propyl alcohol and deionized water in that sequence under sonication), followed by O$_2$ and hexamethyl disilizane (HMDS) priming in order to increase the adhesion between GaP and the subsequent spin-coated electron-beam lithography (EBL) resist of hydrogen silsesquioxane (HSQ). After spin-coating of HSQ layer with a thickness of $\sim$ 540 nm, EBL and development in 25 $\%$ tetra-methyl ammonium hydroxide (TMAH) are carried out to define the nanostructures in HSQ resist. Inductively-coupled plasma reactive ion etching (ICP-RIE) with N$_2$ and Cl$_2$ is then used to transfer the HSQ patterns to the GaP layer. Finally, $\sim$3.2 $\mu$ m SiO$_2$ cladding layer is deposited on top of the structures by ICP-CVD. For optical measurements, an additional layer of polydimethylsiloxane (PDMS) is placed on top of the cladding layer to avoid Fabry-Perot resonances. The structures fabricated are 100 $\mu$m $\times$ 100 $\mu$m in size. 

\subsection*{Experimental setup}
The schematic of the experimental setup is given in the Fig. \ref{FigureS4}. The laser light from an optical parameteric oscillator (OPO) is passed through a dichroic beam splitter cutting-off shorter wavelength components. The pulsed pump is obtained from a Coherent APE OPO tunable from 1100 nm to 1300 nm, pumped by a Ti:Sapphire laser at a wavelength of 830 nm. The CW OPO is a Hubner CWave laser pumped by a 532 nm laser, that is tunable from 900 nm to 1330 nm. The pump light then passes through a variable optical neutral density filter, quarter-wave plate (QWP) and half-wave plate (HWP). The variable neutral density filter is used for varying the pump power levels. The QWP and HWP are jointly used for varying the pump polarization at the input. A polarization extinction of approximately 100 is achieved for the pump light at the input of the sample, as given in Fig. \ref{FigureS4}(b). A polarizer is used for the power measurements and polarization analysis of the input pump. For Fig. 3(b), no polarizer is used at the input. All the other measurement results given in Figs. 1, 2, 3 and 4 use a polarizer at the input of the sample. Lens, L1, is a plano-convex lens with a focal length of 25.4 mm, to achieve a beam waist diameter of approximately 100 um for the pulsed and CW pump. The beam width measurement is carried out using knife edge and the results are given in Fig. \ref{FigureS4}(c). The lens, L2, is an objective with an NA of 0.95. At the output, a flipping mirror is used to image the far-field beam pattern and the sample. A 4f imaging system (L3-L6) with a charged coupled device (CCD) camera is used to image the back-focal plane of the objective, to obtain Fig. 3(d). After this imaging system, there is a filter centered at 650 nm and having a spectral width of 150 nm (Semrock). The output of the filter is directed through a chopper to an amplified photo-detector (APD) to measure the SHG power for CW pump and a power meter (PM) calibrated to the APD is used for the pulsed pump. The $P_{SHG}$ is the average measured SHG power after the filter and $P_{pump}$ is the average measured pump power before L1. The transmission spectrum in Fig. 2(b) is obtained using the same experimental system pumped by a broad band super-continuum source (Leukos) and coupling the light output through the sample, immediately after L2 using a flip mirror, to a high resolution optical spectrum analyzer (OSA, Yokogawa) through a single mode fiber, and taking the ratio of the transmitted signal through the sample with and without the nanostructured array. This fiber coupling system is also used for capturing the spectra of the SHG generated from the metasurface using the spectrometer (Ocean Optics, USB4000).
 \newpage 

\begin{figure}[h]
\centering
\includegraphics[scale=0.6]{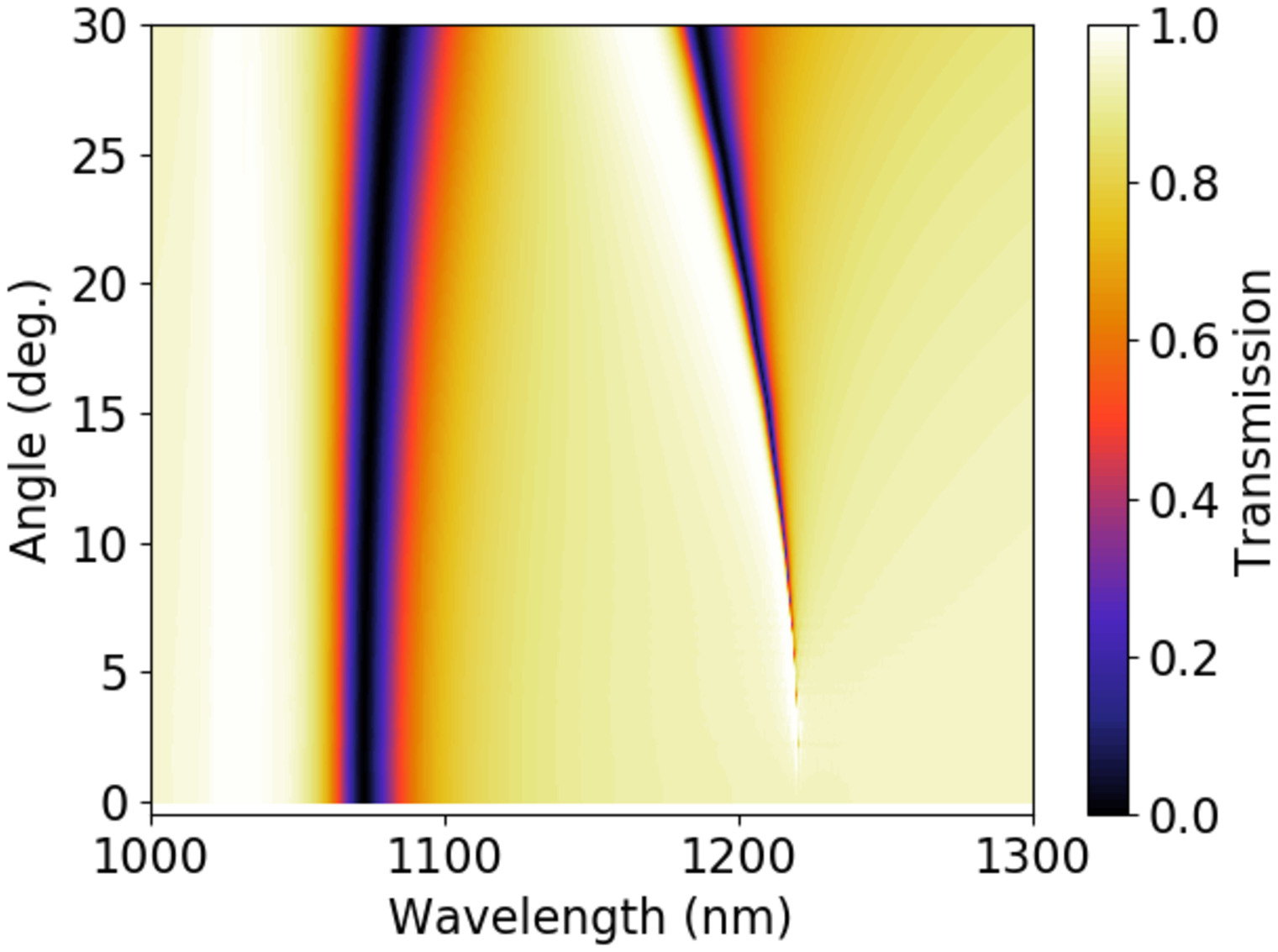}
\caption{Colour map representing the metasurface transmission as a function of wavelength and angle of tilting ($\theta$) of the elliptical cylinders forming the dimers in the metasurface, as shown in Figure 1 in the main text. As seen, the resonance corresponding to the excitation of the quasi-BIC mode experiences a strong narrowing of the linewidth, as well as a decrease in the amplitude, when $\theta$ decreases and the mode evolves towards the perfect BIC. The broad resonance observed at shorter wavelengths, corresponds to the excitation of an in-plane, electric dipole directed along the incident polarization ($p_{x}$), and is almost insensitive to the tilting angle.}
\label{FigureS1}
\end{figure}

\begin{figure}[h]
    \centering
    \includegraphics[scale=0.48]{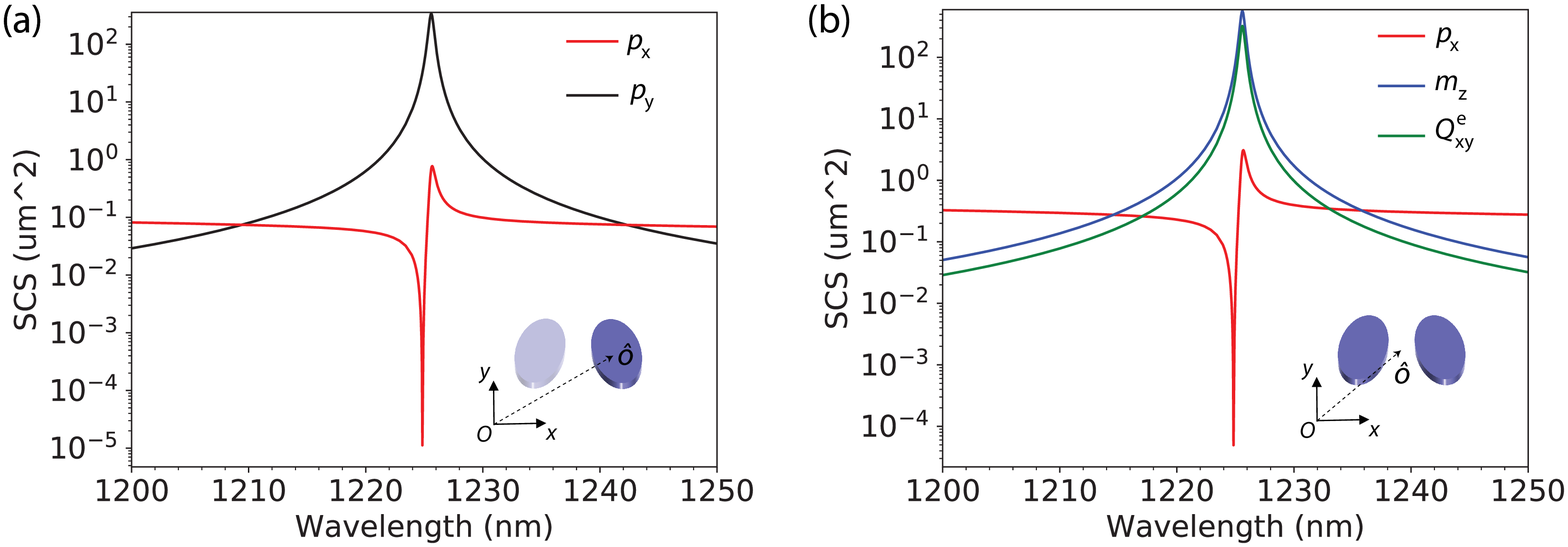}
    \caption{Multipolar decomposition of one unit-cell of the metasurface supporting a quasi-BIC with $\theta = 5^{\circ}$. (a) Non-negligible multipole moments when applying the multipole decomposition, only, to the fields of one of the cylinders comprising the dimer. The origin of coordinates is taken at the center of the cylinder. As can be seen, the only significant multipole moment is a tilted electric dipole moment (as indicated by the presence by both x- and y- components). (b) Non-negligible multipole moments when applying the multipole decomposition to the whole dimer. The origin of coordinates is taken at the center of the dimer. As can be seen, there are three significant multipole moments: an out-of-plane magnetic dipole moment (m$_z$), an in-plane electric quadrupole moment (Q$_{xy}^e$=Q$_{yx}^e$) and, with much lower amplitude, an in-plane electric dipole (with only x-component, double of that of the single cylinder in (a)). The origin of these multipolar components is detailed in the main text, in connection with Fig. 1(b).}
    \label{FigureS2}
\end{figure} 

\begin{figure}[h]
    \centering
    \includegraphics[scale=0.55]{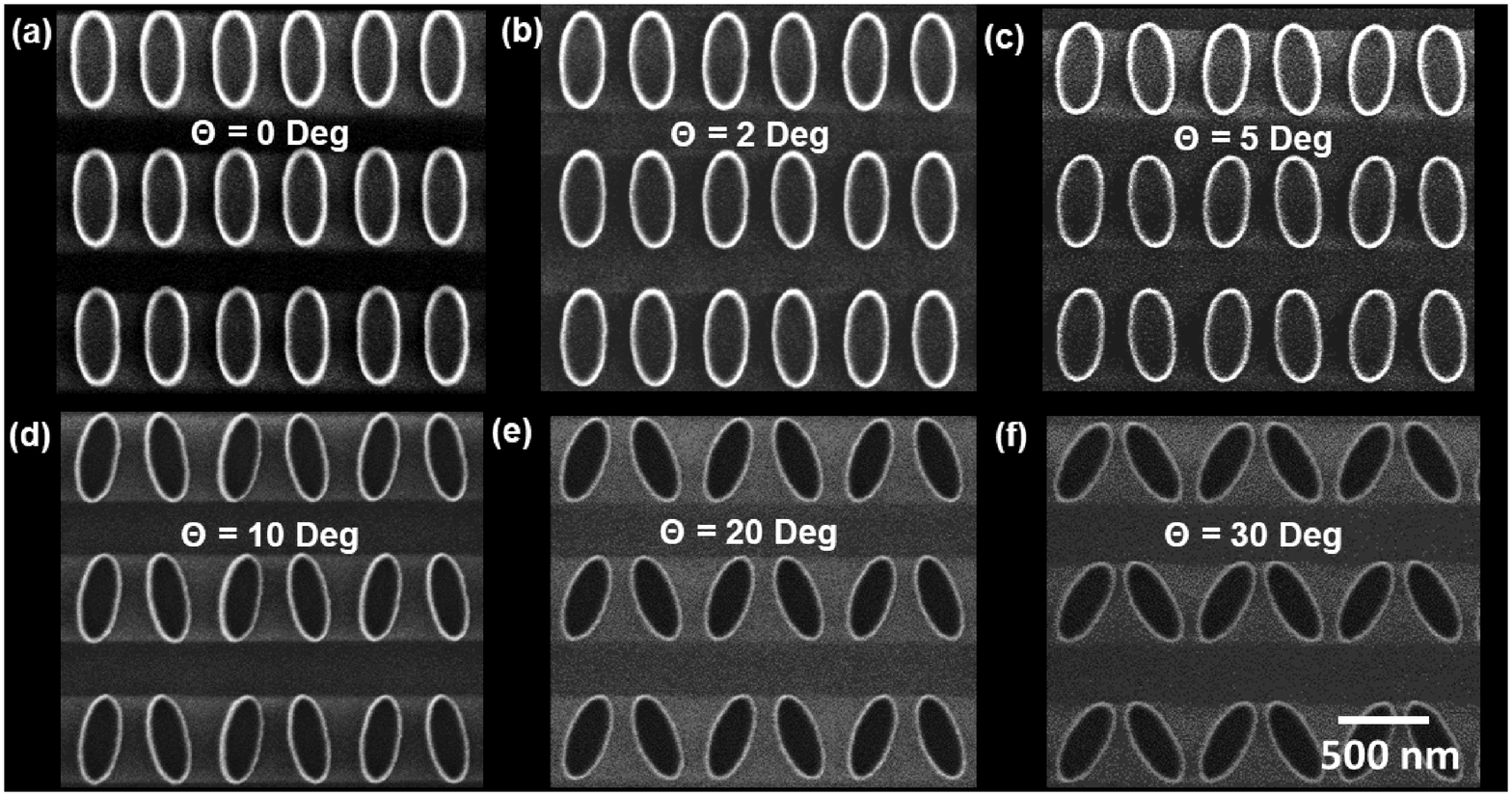}
    \caption{Scanning electron microscope images of the different fabricated metasurfaces studied in this work, comprising elliptical cylinder dimers with relative tilting: (a) $\theta = 0^{\circ}$, (b) $\theta=2^{\circ}$ (for which we were not able to observe the quasi-BIC), (c) $\theta=5^{\circ}$, (d) $\theta = 10^{\circ}$, (e) $\theta = 20^{\circ}$ and (f) $\theta = 30^{\circ}$. The scale bar is the same for all the images.}
    \label{FigureS3}
\end{figure}

\begin{figure}[h]
    \centering
    \includegraphics[scale=0.315]{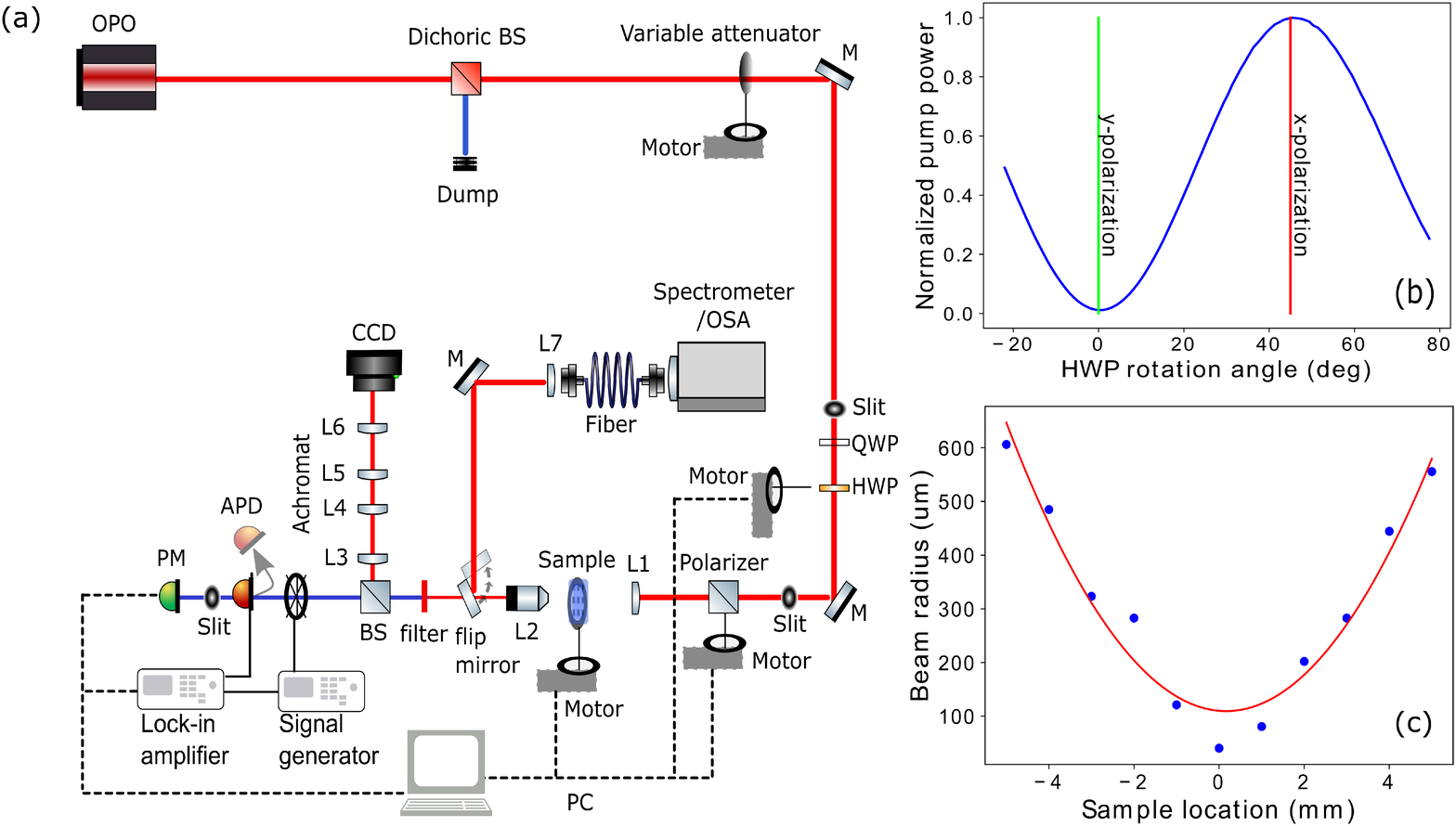}
    \caption{(a) Schematic of the experimental setup used to capture the back focal plane image of the objective, measure the SHG power levels as a function of pump power and pump polarization, and collect the transmission characteristics of the devices. (b) Measured pump power at the output of the polarizer as a function of the HWP rotation angle before L1. (c) Beam radius with respect to the sample position measured using a knife edge.}
    \label{FigureS4}
\end{figure}

\clearpage
\begin{table}[h]
\begin{center}
\caption{Comparison of performances of SHG in dielectric metasurfaces. Efficiency is defined as $P_{SHG}/P_{pump}$, where the power levels are average values.}
\label{tab:Table1}
\begin{tabular}{|p{0.11\textwidth}|p{0.2\textwidth}|p{0.22\textwidth}|p{0.12\textwidth}|p{0.16\textwidth}|}
\hline
\textbf{Platform}	& \textbf{Pump intensity (GW/cm$^2$)}	& \textbf{Pump wavelength (nm)} &  \textbf{Efficiency} &  \textbf{Reference} \\ \hline
\hline
 \textbf{GaP} & \textbf{10e-3} & \textbf{1200} & \textbf{4e-5} & \textbf{This work (Pulse)}  \\
\hline
 \textbf{GaP} & \textbf{1e-6} & \textbf{1200} & \textbf{2e-7} & \textbf{This work (CW)} \\
\hline
 AlGaAs & 0.22 & 1570 & 1.2e-5 & \cite{Koshelev2020} \\
\hline
 AlGaAs & 7 & 1550 & 8e-6 & \cite{Kruk2017a}\\
\hline
 AlGaAs & 1.72 & 1556 & 9e-6 & \cite{Gili2016, Ghirardini2017} \\
\hline
 GaP & 100 & 910 - 1300 & 2e-6 & \cite{Cambiasso2017}\\
\hline
 GaAs & 0.32e-3 & 1000 & 6e-6 & \cite{Vabishchevich2018} \\
\hline
 GaAs & 3.4 & 1020 & 2e-5 & \cite{Liu2016} \\
\hline
 ZnO & 15 & 394 & 1e-8 & \cite{Semmlinger2018}\\
\hline

\end{tabular}
\end{center}
\end{table}

\bibliography{SHG_BIC}

\end{document}